\documentclass[12pt]{article}
\usepackage{amsmath, amssymb}
\begin{document}
\date{}

\title{The $su(1,1)$ dynamical algebra from the Schr\"odinger ladder operators for $N$-dimensional systems: hydrogen atom, Mie-type potential, harmonic oscillator and pseudo-harmonic oscillator.}

\author{D. Mart\'{\i}nez$^{a}$\footnote{{\it E-mail address:} dmartinezs77@yahoo.com.mx}, J. C. Flores-Urbina$^{b}$\\ R. D. Mota$^{b}$ and V. D. Granados$^{c}$} \maketitle

\begin{minipage}{0.9\textwidth}
\small $^{a}$ Universidad Aut\'onoma de la Ciudad de M\'exico,
Plantel Cuautepec, Av. La Corona 320, Col. Loma la Palma,
Delegaci\'on
Gustavo A. Madero, 07160, M\'exico D. F., M\'exico.\\

\small $^{b}$ Unidad Profesional Interdisciplinaria de Ingenier\'{\i}a y
Tecnolog\'{\i}as Avanzadas, IPN. Av. Instituto Polit\'ecnico
Nacional 2580, Col. La Laguna Ticom\'an, Delegaci\'on Gustavo A.
Madero, 07340 M\'exico D. F., M\'exico.\\

\small $^{c}$Escuela Superior de F\'{\i}sica y Matem\'aticas,
Instituto Polit\'ecnico Nacional,
Ed. 9, Unidad Profesional Adolfo L\'opez Mateos, 07738 M\'exico D F, M\'exico.
\end{minipage}

\begin{abstract}
We apply the Schr\"odinger factorization to construct the ladder operators for hydrogen atom, Mie-type potential, harmonic oscillator and pseudo-harmonic oscillator in arbitrary dimensions. By generalizing these operators we show that the dynamical algebra for these problems is the $su(1,1)$ Lie algebra.
\end{abstract}
\section{Introduction}
The factorization methods have played an important role in the
study of quantum systems \cite{DIRAC}-\cite{INFELD-HULL}. This is because, if the Schr\"odinger equation is factorizable,
the energy spectrum and the eigenfunctions are obtained algebraically. Infeld and
Hull \cite{INFELD,INFELD-HULL} used the ideas of Dirac \cite{DIRAC} and Schr\"odinger \cite{SCH1}-\cite{SCH3} and created a factorization
method (IHFM) which uses a particular solution for the Ricatti equation. This
method allowed to classify the problems according to the characteristics involved in the
potentials and it is closely related to the supersymmetric quantum mechanics \cite{WITTEN}-\cite{LAHIRI}.

The Schr\"odinger factorization  developed in reference \cite{SCH1} 
is a technique which is essentially different to that of the IHFM, as it was pointed out by Infeld \cite{INFELD}: "Schr\"odinger uses a finite number of infinite ladders, whereas I use an infinite number of finite ladders''. In order to clarify these differences we consider a typical central-potential problem like the two-dimensional hydrogen atom  or the two-dimensional isotropic harmonic oscillator. Their dynamical algebras for the bound states are compact. Therefore the corresponding ladder operators can only generate a finite number of states with different values of the angular momentum for each degenerate energy eigenvalue \cite{CORDERO}. On the other hand, their dynamical  algebras are non-compact and have infinite-dimensional representations \cite{CORDERO}. Thus they generate an infinite number of states with different values of the energy for each degenerate angular momentum eigenvalue. Historically the Schr\"odinger factorization has been less applied to physical problems than the IHFM. 

A systematic method to find both compact and non-compact
algebra generators for a given system has not still developed. These generators have been
intuitively found and forced them to close an algebra, as it is extensively shown in reference
\cite{BARUT, COOPER}.
For the most important central-potential problems, the hydrogen atom and the isotropic
harmonic oscillator, algebraic treatments by means of compact and non-compact
groups are known \cite{WYBOURNE,ENGLEFIELD}. Recently, the generators of these groups for some two-dimensional central-potential problems were obtained by means of factorization methods \cite{DANIEL} and it has been shown that the Schr\"odinger factorization operators are related to the dynamical algebra generators for these systems \cite{DANIEL}.

In the last two decades the $N$-dimensional systems have been treated by many authors \cite{BAGCHI2}. The study of these systems is important because mathematically some few-body problems in three dimensions are equivalent to higher-dimensional ones \cite{SAELEN}. It is well known that the $N$-dimensional problems of hydrogen atom and harmonic oscillator are exactly solvable and accept to be studied by supersymmetric quantum mechanics \cite{NIETO,LAHIRI2,NEGRO}. Recently, it has been shown that the $N$-dimensional Mie-type potential \cite{IKHDAIR, AGBOOLA} and the pseudo-harmonic oscillator \cite{OYEWUMI} are exactly solvable potentials. In these works, by using the recursion relations for the generalized Laguerre polynomials and the explicit form of the eigenfunctions, the $su(1,1)$ dynamical algebra generators for these systems are found.

The aim of this work is to obtain the generators of the $su(1,1)$ dynamical algebra from the Schr\"odinger factorization for the hydrogen atom, Mie-type potential, harmonic oscillator and pseudo-harmonic oscillator in arbitrary dimensions. This paper is organized as follows. Section 2 concerns with the harmonic oscillator and the hydrogen atom. Section 3 is dedicated to the study of the Mie-potential and the pseudo-harmonic oscillator. In Section 4 we give the concluding remarks.

\section{Harmonic Oscillator and Hydrogen Atom}

\subsection{Radial Schr\"odinger Equation in $N$ Dimensions}

The $N$-dimensional Schr\"odinger equation for bound states
\begin{equation}
\left(-\frac{\hbar^2}{2m}\Delta_N+V(r)\right)\Psi(r,\Omega_N)=E\Psi(r,\Omega_N)
\end{equation}
can be reduced to the radial equation \cite{BAGCHI2,SAELEN}
\begin{equation}
\left(\frac{d^2}{dr^2} + \frac{N-1}{r}\frac{d}{dr}-\frac{\ell(\ell+N-2)}{r^2}-\frac{2m}{\hbar^2}V(r)
+ \frac{2m}{\hbar^2}E_n\right)R_{n\ell}(r)=0,\label{radial1}
\end{equation}
where $\Delta_N=\frac{\partial^2}{\partial x_1^2}+...+\frac{\partial^2}{\partial x_N^2}$,   $\Psi(r,\Omega_N)=\Psi_{n\ell m}(r,\Omega_N)$$=R_{n\ell}(r)Y_\ell^m(\Omega_N)$ and $Y_\ell^m(\Omega_N)$ are the hyperspherical harmonics.
If we propose $R_{n\ell}(r)$ to have the form 
\begin{equation}
R_{n\ell}(r)=r^{\frac{1-N}{2}}U_{n\ell}(r),
\label{ru}
\end{equation}
then, equation (\ref{radial1}) can be rewritten as
\begin{equation}
\left( r^2\frac{d^2}{dr^2}-\frac{2m}{\hbar^2}\left(V(r)-E_n\right)r^2 -\frac{(N-1)(N-3)}{4}-\ell(\ell+N-2)\right) U_{n\ell}=0.
\label{SE}
\end{equation}
This expression is important to obtain the main results of this paper.

\subsection{Harmonic Oscillator}
By substituting the harmonic oscillator potential  $V(r)=\frac{m\omega^2r^2}{2}$ into equation (\ref{SE}) we have 
\begin{equation}
\left \lbrace r^2\frac{d^2}{dr^2}-\frac{2m}{\hbar^2}\left(\frac{m\omega^2r^2}{2}-E_n\right)r^2 -\frac{(N-1)(N-3)}{4}-\ell(\ell+N-2)\right\rbrace U_{n\ell}=0.
\end{equation}
If we set $r=\alpha^{-1} x$, $\alpha=\sqrt{m\omega/\hbar}$ and $\lambda_n=\frac{E_n}{\hbar \omega}$, this equation can be written as follows
\begin{equation}
\left(x^2\frac{d^2}{dx^2}-x^4+2\lambda_n x^2\right)U_{n\ell}=\left(\ell(\ell+N-2)+\frac{(N-1)(N-3)}{4}\right)U_{n\ell}.
\label{seis}
\end{equation}
Since the right hand side of this equation can be factorized as   
\begin{equation}
\ell(\ell+N-2)+\frac{(N-1)(N-3)}{4}=\left(\ell+\frac{N-2}{2}\right)^2-\frac{1}{4} \equiv \kappa^2-\frac{1}{4},
\label{kappa}
\end{equation}
equation (\ref{seis}) is rewritten as 
\begin{equation}
L_n U_{n\ell}\equiv \left(-x^2\frac{d^2}{dx^2}+x^4-2\lambda_n x^2\right)U_{n\ell}=-\left(\kappa^2-\frac{1}{4}\right)U_{n\ell}.
\label{ln}
\end{equation}

In order to factorize the operator $L_n$ we apply the Schr\"odinger factorization \cite{SCH1,DANIEL}. Thus, we propose a pair of first-order differential operators such that  
\begin{equation}
\left(x\frac{d}{dx}+ax^2+b\right)\left(-x\frac{d}{dx}+cx^2+f\right)U_{n\ell}=gU_{n\ell},
\label{sch}
\end{equation}
where  $a$, $b$, $c$, $f$ y $g$ are constants to be determined. Expanding this expression and comparing it with equation (\ref{ln}) we obtain
\begin{equation}
a=c=\mp1,\hspace{2ex} f=b+1=-\frac{1}{2}\pm \lambda_n,\hspace{2ex} g=f(f-1)-\left(\kappa^2-\frac{1}{4}\right).
\end{equation}
Using these results, equation (\ref{ln}) is equivalent to 
\begin{eqnarray}
(D_-^n-1)D_+^{n} U_{n\ell}=\frac{1}{4}\left[\left(\lambda_n+\frac{1}{2}\right)\left(\lambda_n+\frac{3}{2}\right)-\left(\kappa^2-\frac{1}{4}\right)\right]U_{n\ell},\label{dn1}\\
(D_+^n+1)D_-^{n} U_{n\ell}=\frac{1}{4}\left[\left(\lambda_n-\frac{1}{2}\right)\left(\lambda_n-\frac{3}{2}\right)-\left(\kappa^2-\frac{1}{4}\right)\right]U_{n\ell}\label{dn2},
\end{eqnarray}
where we have defined the operators
\begin{equation}
D_{\pm}^n=\frac{1}{2}\left(\mp
x\frac{d}{dx}+x^2-\lambda_n\mp\frac{1}{2}\right). \label{Dn}
\end{equation}
  
To find the dynamical algebra we define the operator 
\begin{equation}
D_3=\frac{1}{4}\left(-\frac{d^2}{dx^2}+x^2+\frac{\kappa^2-\frac{1}{4}}{x^2}\right),
\label{D3}
\end{equation}
which from equation (\ref{ln}) satisfy 
\begin{equation}
D_3U_{n\ell}=\frac{\lambda_n}{2}U_{n\ell} \label{D3a}.
\end{equation}

In order to generalize the operators $D^n_{\pm}$ to $E_n$-independent ones, we use equation (\ref{D3}) and obtain 
\begin{equation}
D_{\pm}=\frac{1}{2}\left(\mp x\frac{d}{dx}+x^2-2D_3\mp\frac{1}{2}\right). \label{D+-}
\end{equation}

The properties of the operators $D_3$ and $D_{\pm}$ are obtained by considering their explicit form and the inner product for the radial eigenfunctions of the Schr\"odinger equation (\ref{radial1}) on the Hilbert space defined by
\begin{equation}
(R_{n'\ell}\:,R_{n\ell})\equiv\int_0^\infty R^{*}_{n'\ell}(r)R_{n\ell}(r)r^{N-1}dr.\label{innerosc}
\end{equation}
From equations (\ref{ru}) and (\ref{innerosc}), we find that the inner product for the functions $U_{n\ell}$ is
\begin{equation}
(U_{n'\ell}\:,U_{n\ell})\equiv\int_0^\infty U^{*}_{n'\ell}(x)U_{n\ell}(x)dx.\label{ortosc}
\end{equation}

Since equation (\ref{D3a}) is essentially the Schr\"odinger equation, we can immediately prove that  $D_3$ is hermitian. Moreover, using equation (\ref{D+-}) and (\ref{ortosc}) we can show
\begin{equation}
(U_{n'\ell}\:,D_{\pm}U_{n\ell})=(D_{\mp}U_{n'\ell}\:,U_{n\ell}),
\end{equation}
which implies $D_{\pm}=D_{\mp}^\dagger$. This means that the operators $D_\pm$ are mutually adjoint.  

By direct calculation we show that the operators $D_\pm$ and $D_3$ satisfy the $su(1,1)$ Lie algebra
\begin{align}
[D_\pm,D_3]& =\mp D_\pm,\\
\lbrack D_+,D_-\rbrack& =-2D_3,
\end{align}
with Casimir operator 
\begin{equation}
D^2\equiv -D_{\pm}D_{\mp}+D_3^2\mp D_3.
\end{equation}
Explicitly from equations (\ref{D3}) and (\ref{D+-}), the eigenvalue equation for this operator is
\begin{align}
D^2 U_{n\ell}&=\left(\frac{\kappa^2-1}{4}\right)U_{n\ell}.\label{Cas}
\end{align}
According to the theory of representations \cite{ADAMS2} and equations (\ref{D3a}) and (\ref{Cas}), we find that the action of the raising and lowering operators on the eigenstates $U_{n\ell}$ is 
\begin{align}
D_{+} U_{n\ell}&=\frac{1}{2}\sqrt{(\lambda_n-\kappa+1)(\lambda_n+\kappa+1)}\: U_{n+1\:\ell},\label{d+}\\
D_{-} U_{n\ell}&=\frac{1}{2}\sqrt{(\lambda_n-\kappa-1)(\lambda_n+\kappa-1)}\: U_{n-1\:\ell}\label{d-}.
\end{align} 
This means that the generators of the $su(1,1)$ algebra are represented by infinite-dimensional subspaces of the radial quantum states. Moreover, the action of the Schr\"odinger operators on the eigenstates $U_{n\ell}$ is to change only the radial quantum number $n$ in one unit. Nevertheless, this action implies a change in two units for the principal quantum number $\nu\equiv2n+\ell$.

\subsection{Hydrogen Atom}
Substituting the Coulomb potential  $V(r)=-\frac{e^2}{r}$ into equation (\ref{SE}) we have 
\begin{equation}
\left \lbrace r^2\frac{d^2}{dr^2}+\frac{2m}{\hbar^2}\left(\frac{e^2}{r}+E_n\right)r^2 -\frac{(N-1)(N-3)}{4}-\ell(\ell+N-2)\right\rbrace\tilde U_{n\ell}=0.
\label{ah}
\end{equation}
We set $\xi=\frac{me^2}{\hbar^2}$, $K^2_n=-\frac{me^4}{2\hbar^2E_n}$ and $r=K_nx$ to rewrite equation (\ref{ah}) as
\begin{equation}
\tilde{L}_n\tilde U_{n\ell}\equiv \left(-x^2\frac{d^2}{dx^2}-2\xi K_nx+{\xi}^2 x^2\right)\tilde U_{n\ell}=-\left(\kappa^2-\frac{1}{4}\right)\tilde U_{n\ell}.
\label{Ln}
\end{equation}
An analogous procedure to that followed for the case of the $N$-dimensional harmonic oscillator allows us to find the Schr\"odinger operators for the $N$-dimensional hydrogen atom 
\begin{equation}
T^n_\pm=\mp x\frac{d}{dx}+\xi x-K_n,
\label{Tn}
\end{equation}
which satisfy 
\begin{eqnarray}
(T_-^n-1)T_+^n \tilde U_{n\ell}=\left[K_n(K_n+1)-\left(\kappa^2-\frac{1}{4}\right)\right]\tilde U_{n\ell},\label{tn1}\\
(T_+^n+1)T_-^n \tilde U_{n\ell}=\left[K_n(K_n-1)-\left(\kappa^2-\frac{1}{4}\right)\right]\tilde U_{n\ell}.\label{tn2}
\end{eqnarray}

Equation (\ref{Ln}) allows to define the operator 
\begin{equation}
T_3=\frac{1}{2\xi}\left(-x\frac{d^2}{dx^2}+\xi^2x+\frac{\kappa^2-\frac{1}{4}}{x}\right),
\label{T3}
\end{equation}
which satisfies
\begin{equation}
T_3\tilde U_{n\ell}=K_n\tilde U_{n\ell}.\label{T3eigen}
\end{equation}

Thus, from equation (\ref{Tn}) and (\ref{T3}) we obtain the operators  
\begin{equation}
T_{\pm}=\mp
x\frac{d}{dx}+\xi x-T_3. \label{T+-}
\end{equation}
Note that these operators depend implicitly (via $r=K_n x$) on the value of $n$ and act on the eigenstates which are characterized by this value of $n$.

If we apply the inner product satisfied by the eigenfunctions of the harmonic oscillator, equation (\ref{ortosc}), we find that the operator $T_3$ is not hermitian with respect to this scalar product \cite{ADAMS2,ADAMS}. Nevertheless, if we define the new inner product on the Hilbert space spanned by the radial eigenfunctions of the $N$-dimensional hydrogen atom as 
\begin{equation}
(\tilde R_{n'\ell}\:,\tilde R_{n\ell})\equiv\int_0^\infty \tilde R^{*}_{n'\ell}(r)\tilde R_{n\ell}(r)r^{N-2}dr\label{innerat}
\end{equation}
or equivalently
\begin{equation}
(\tilde U_{n'\ell}\:,\tilde U_{n\ell})\equiv\int_0^\infty \tilde U^{*}_{n'\ell}(x)\tilde U_{n\ell}(x)x^{-1}dx,\label{ortat}
\end{equation}
we show that the operator $T_3$ is hermitian with respect to this scalar product. Moreover, using (\ref{T+-}) and (\ref{ortat}) we show
\begin{equation}
(\tilde U_{n'\ell}\:,T_{\pm}\tilde U_{n\ell})=(T_{\mp}\tilde U_{n'\ell}\:,\tilde U_{n\ell}),
\end{equation}
which implies that $T_{\pm}=T_{\mp}^\dagger$.

On the other hand, by direct calculation it is immediate to show that the commutation relations satisfied by the operators $T_{\pm}$ and $T_{3}$ are
\begin{eqnarray}
[T_\pm,T_3]& =\mp T_\pm,\\
\lbrack T_+,T_-\rbrack& =-2T_3,
\end{eqnarray}
which is the $su(1,1)$ Lie algebra. By using equations (\ref{T3}) and (\ref{T+-}) we find that the eigenvalue equation for the corresponding Casimir operator is
\begin{equation}
T^2 \tilde U_{n\ell}\equiv\left(-T_{\pm}T_{\mp}+T_3^2\mp T_3\right)\tilde U_{n\ell}=\left(\kappa^2-\frac{1}{4}\right)\tilde U_{n\ell}\label{CasT}.
\end{equation}
Hence, from the theory of representations \cite{ADAMS2} and equations (\ref{T3eigen}) and (\ref{CasT}), for the ladder operators we have
\begin{align}
T_{+}\tilde U_{n\ell}&=\sqrt{(K_n-\kappa+\frac{1}{2})(K_n+\kappa+\frac{1}{2})}\:\tilde U_{n+1\:\ell},\label{t+}\\
T_{-}\tilde U_{n\ell}&=\sqrt{(K_n-\kappa-\frac{1}{2})(K_n+\kappa-\frac{1}{2})}\:\tilde U_{n-1\:\ell}.\label{t-}
\end{align}

In this way, we have showed that from the Schr\"odinger factorization it is possible to find operators which relate eigenstates belonging to different energy and same angular momentum. Moreover, this factorization allowed us to construct the generators of the $su(1,1)$ Lie algebra for each Hamiltonian and to obtain the corresponding recurrence relations (\ref{d+}) and (\ref{d-})  or (\ref{t+}) and (\ref{t-}). In addition to this, the generators $D_{\pm}$ and $T_{\pm}$ are represented by infinite-dimensional Hilbert subspaces of the radial quantum states.

The operators given in (\ref{T3}) and (\ref{T+-}), written in terms of the variable $r$ and acting on the radial functions $\tilde R_{n\ell}(r)$ are expressed as
\begin{align}
\tau_3&\equiv\frac{K_n}{2\xi}\left[-r\frac{d^2}{dr^2}-(N-1)\frac{d}{dr}+\left(\frac{\xi}{K_n}\right)^2r+\frac{\kappa^2-\left(\frac{N-2}{2}\right)^{2}}{r}\right],\\
\tau_{\pm}&\equiv\mp r\frac{d}{dr}+\frac{\xi}{K_n}r\mp\frac{N-1}{2}-\tau_3.
\end{align}
By direct calculation we show that this set of operators satisfy the $su(1,1)$ Lie algebra, as it was expected.

Similarly, for the harmonic oscillator we show that the operators $D_3$ and $D_{\pm}$, equations (\ref{D3}) and (\ref{D+-}), rewritten in terms of the variable $r$ and acting on the functions $R_{n\ell}(r)$ satisfy the $su(1,1)$ algebra.

\section{Pseudo-Harmonic Oscillator and Mie-Type Potential}
\subsection{Pseudo-harmonic potential}
The pseudo-harmonic oscillator potential is given by 
\begin{equation}
V(r)=Ar^2+\frac{B}{r^2}+C.
\end{equation}  
With this potential and the definitions 
\begin{equation}
x=r\alpha,\hspace{5ex}\Lambda_n=\frac{\epsilon_n}{\hbar\omega},\hspace{5ex}\epsilon_n=E_n-C,
\end{equation}
equation (\ref{SE}) can be written as
\begin{align}
\left(-x^2\frac{d^2}{dx^2}+x^4-2\Lambda_n x^2\right){\cal U}_{n\ell}=&\left(\frac{(N-1)(N-3)}{4}+\ell(\ell+N-2)+\frac{2mB}{\hbar^2}\right){\cal U}_{n\ell},\nonumber\\
=&-\left(\beta^2-\frac{1}{4}\right){\cal U}_{n\ell},
\label{clave}
\end{align}
where 
\begin{equation}
\beta=\sqrt{(2\ell+N-2)^2+\frac{8mB}{\hbar^2}}.
\end{equation}
Since equation (\ref{clave}) is formally equal to that 
for the harmonic oscillator, it is immediate to find and generalize the corresponding Schr\"odinger factorization operators. Therefore, by making $\kappa\rightarrow\beta$ in equations (\ref{D3}), (\ref{D+-}) and (\ref{Cas}) we obtain the set of operators
\begin{align}
{\mathcal D}_3& =\frac{1}{4}\left(-\frac{d^2}{dx^2}+x^2+\frac{\beta^2-\frac{1}{4}}{x^2}\right),\label{dd3}\\
{\mathcal D}_{\pm}& =\frac{1}{2}\left(\mp x\frac{d}{dx}+x^2-2{\mathcal D}_3\mp\frac{1}{2}\right),
\end{align}   
which close the $su(1,1)$ dynamical algebra for the pseudo-harmonic oscillator
\begin{eqnarray}
[{\mathcal D}_\pm,{\mathcal D}_3]& =\mp {\mathcal D}_\pm,\label{D}\\
\lbrack {\mathcal D}_+,{\mathcal D}_-\rbrack& =-2{\mathcal D}_3.
\end{eqnarray}
The corresponding Casimir operator satisfies
\begin{equation}
{\mathcal D}^2 U_{n\ell}=\left(\frac{\beta^2-1}{4}\right){\mathcal U}_{n\ell}.
\end{equation}

If we make $\lambda_n\rightarrow\Lambda_n$ in equations (\ref{D3a}), (\ref{d+}) and (\ref{d-}) then, the action of the generators ${\mathcal D}_{\pm}$ and ${\mathcal D}_3$ on the eigenstates of the pseudo-harmonic oscillator is
\begin{align}
{\mathcal D}_{\pm} {\mathcal U}_{n\ell}&=\frac{1}{2}\sqrt{(\Lambda_n-\beta\pm1)(\Lambda_n+\beta\pm1)}\: {\mathcal U}_{n\pm1\:\ell},\\
{\mathcal D}_3 {\mathcal U}_{n\ell}&=\frac{\Lambda_n}{2}{\mathcal U}_{n\ell}.
\end{align}
A similar procedure to that followed in section 2.1 allows to show that  
${\mathcal D}_3={\mathcal D}_3^\dag$ and ${\mathcal D}_\pm={\mathcal D}_\mp^\dag$.

\subsection{Mie-type potential}

This section is advocated to obtain the dynamical algebra for the Mie-type potential given by 
\begin{equation}
V(r)=\frac{A'}{r}+\frac{B'}{r^2}+C'.
\end{equation}  
This potential and the definitions 
\begin{equation}
r=\Sigma_nx,\hspace{5ex}\Sigma_n^2=-\frac{m(A')^2}{2\hbar^2\tilde\epsilon_n},\hspace{5ex}\zeta=-\frac{mA'}{\hbar^2},\hspace{5ex}\tilde\epsilon_n=E_n-C',
\end{equation}
allow us to rewrite equation (\ref{SE}) as
\begin{align}
\left(-x^2\frac{d^2}{dx^2}-2\zeta\Sigma_n x+ \zeta^2x^2\right){\tilde{\cal U}}_{n\ell}& =\left(\frac{(N-1)(N-3)}{4}+\ell(\ell+N-2)+\frac{2mB'}{\hbar^2}\right){\tilde{\cal U}}_{n\ell},\nonumber\\
& =-\left(\gamma^2-\frac{1}{4}\right){\tilde{\cal U}}_{n\ell},
\label{clave1}
\end{align}
where 
\begin{equation}
\gamma=\sqrt{(2\ell+N-2)^2+\frac{8mB'}{\hbar^2}}.
\end{equation}
Since this equation is formally equal to that for the hydrogen atom, the generators for the $N$-dimensional Mie-type potential are
\begin{align}
{\mathcal T}_3& =\frac{1}{2\zeta}\left(-x\frac{d^2}{dx^2}+\zeta^2x+\frac{\gamma^2-\frac{1}{4}}{x}\right),
\label{MIET3}\\
{\mathcal T}_{\pm}& =\mp x\frac{d}{dx}+\zeta x-{\mathcal T}_3. \label{MIET+-}
\end{align}
We can show that these operators satisfy the $su(1,1)$ Lie algebra
\begin{eqnarray}
[{\mathcal T}_\pm,{\mathcal T}_3]& =\mp {\mathcal T}_\pm,\label{T}\\
\lbrack {\mathcal T}_+,{\mathcal T}_-\rbrack& =-2{\mathcal T}_3
\end{eqnarray}
and similarly to section 2.2, these operators have the properties  ${\mathcal T}_3={\mathcal T}_3^\dag$ and ${\mathcal T}_\pm ={\mathcal T}_\mp^\dag$.

In addition to this, the eigenvalue equation for the Casimir operator and the action of the generators ${\mathcal T}_{\pm}$ and ${\mathcal T}_3$ on the eigenstates $\tilde{\mathcal U}_{n\ell}$ are
\begin{equation}
{\mathcal T}^2 \tilde{\mathcal U}_{n\ell}=\left(\gamma^2-\frac{1}{4}\right)\tilde{\mathcal U}_{n\ell}
\end{equation}
and 
\begin{align}
{\mathcal T}_{\pm}\tilde{\mathcal U}_{n\ell}&=\sqrt{(\Sigma_n-\gamma\pm\frac{1}{2})(\Sigma_n+\gamma\pm\frac{1}{2})}\:\tilde {\mathcal U}_{n+1\:\ell},\\
{\mathcal T}_3 \tilde{\mathcal U}_{n\ell}&=\Sigma_n\tilde{\mathcal U}_{n\ell},
\end{align}
respectively.

By transforming the pseudo-harmonic and Mie-type potential to the harmonic oscillator and the hydrogen atom, respectively, we were capable to find the corresponding generators which close the $su(1,1)$ dynamical symmetry in a simple way. From the commutation relation (\ref{D}), we obtain the recurrence relations for the pseudo-harmonic oscillator ${\cal D}_\pm{\cal U}_{n\ell}\propto{\cal U}_{n\pm1\;\ell}$, meanwhile from (\ref{T}) we obtain the Mie-type potential recurrence relations ${\cal T}_\pm{\tilde{\cal U}}_{n\ell}\propto{\tilde{\cal U}}_{n\pm1\;\ell}$. It must be pointed out that we achieved these results without the use of the explicit form of the eigenfunctions.

\section{Concluding Remarks}
 
We studied the $N$-dimensional harmonic oscillator and the hydrogen atom in an integrated approach by applying the Schr\"odinger factorization. In this way a pair of first-order differential operators were obtained whose action on the eigenstates of the corresponding Hamiltonian is to change only the radial quantum number $n$ leaving fixed the quantum angular momentum number $\ell$. From the corresponding Hamiltonian we introduced, in a natural way, a third differential operator which closes the $su(1,1)$ algebra for these problems. In order to find the dynamical algebra of the $N$-dimensional Mie-type potential and the pseudo-harmonic oscillator we reduced the corresponding radial Schr\"odinger equations to those of the hydrogen atom and the harmonic oscillator, respectively. In all the problems studied in this work, we found that the generators of the $su(1,1)$ algebra  are represented by infinite-dimensional Hilbert subspaces of the radial quantum states. We emphasize that in any calculation we did not use the explicit form of the eigenfunctions.

It must be noticed that the method we applied to construct the generators of the algebra $su(1,1)$ differs from those followed by other authors \cite{NEGRO,AGBOOLA,OYEWUMI,GUR}. For instance, in  \cite{NEGRO} to obtain the dynamical algebra generators for the harmonic oscillator the SUSY operators were used, meanwhile in this paper we constructed them by applying the Schr\"odinger factorization. On the other hand,  the authors of the  references \cite{AGBOOLA, OYEWUMI} used the recursion relations for the generalized Laguerre polynomials and the explicit expressions of the radial eigenfunctions to find the differential form of only two operators. Even though they close the  $su(1,1)$ dynamical algebra, the third "operator" they used is not an operator but the eigenvalue of the operator itself.

The Schr\"odinger factorization can be applied successfully to study more complex systems such as matrix potentials with axial symmetry and generalized MICZ-Kepler problem which is a work in progress.
\section*{Acknowledgments}

This work was partially supported by SNI-M\'exico, CONACYT grant
number J1-60621-I, COFAA-IPN, EDI-IPN, SIP-IPN projects numbers
20091042 and 20090590, and ADI-UACM project number 7D92123004.


\begin{thebibliography}{99}
\bibitem{DIRAC}P. A. M. Dirac, The Principles of Quantum Mechanics, Clarendon Press, Oxford, 1935.
\bibitem{SCH1}E. Schr\"odinger, Proc. R. Ir. Acad. A 46(1940) 9.
\bibitem{SCH2}E. Schr\"odinger, Proc. R. Ir. Acad. A 46(1940)183.
\bibitem{SCH3}E. Schr\"odinger, Proc. R. Ir. Acad. A 47(1941)53.
\bibitem{INFELD}L. Infeld, Phys. Rev. 59(1941)737.
\bibitem{INFELD-HULL}L. Infeld  and T. E. Hull, Rev. Mod. Phys. 23(1951)21.
\bibitem{WITTEN}E. Witten, Nucl. Phys. B 188(1981)513.
\bibitem{COOPER1} F. Cooper, A. Khare and U. Sukhatme, Phys. Rep. 251(1995)267-385.
\bibitem{COOPER2}R. Dutt, A. Khare, U.P. Sukatme, Am. J. Phys. 56(1987)163.
\bibitem{BAGCHI2}B. K. Bagchi, Supersymmetry in Quantum and Classical Mechanics,
Chapman \& Hall/CRC, USA, 2001.
\bibitem{SAELEN}L. Saelen, R. Nepstad, J. P. Hansen and L. B. Madsen, J. Phys. A. Math. Theor. 40(1097)2007.
\bibitem{LAHIRI}A. Lahiri, P. Roy and B. Bagchi, Int. J. Mod. Phys. A
5(1990)1383-1456.
\bibitem{CORDERO}P. Cordero and J. Daboul, J. Math.  Phys. 46(2005)053507.
\bibitem{BARUT}A. Bohm, Y. Ne'eman, A. O. Barut, Dynamical Groups and Spectrum Generating Algebras, vols. 1 and 2,
World Scientific, Singapore, 1988.
\bibitem{COOPER} I. L. Cooper, J. Phys. A: Math Gen. 26(1993)1601.
\bibitem{WYBOURNE}B. G. Wybourne, Classical Groups for Physicists, Wiley Interscience, New York, 1974.
\bibitem{ENGLEFIELD}M. J. Englefield, Group Theory and the Coulomb Problem, Wiley Interscience, USA, 1972.
\bibitem{DANIEL}D. Mart\'{\i}nez, R. D. Mota, Ann. Phys. 323(2008)1024.
\bibitem{NIETO}V. A. Kosteleck\'y, M. M. Nieto and D. R. Truax, Phys. Rev. D32 (1985)2627.
\bibitem{LAHIRI2}A. Lahiri, P.K. Roy and B. Bagchi, Phys. Rev. A 38(1988)1989.
\bibitem{NEGRO}D. J. Fern\'andez, J. Negro and M. A. del Olmo, Ann. Phys. 252(1996)386.
\bibitem{IKHDAIR}S. Ikhdair and R. Sever, J. Mol. Struc. (Theochem) 13(2008)855.
\bibitem{AGBOOLA}D. Agboola., arXiv:0812.3780v3.
\bibitem{OYEWUMI}K. J. Oyewumi, F. O. Akinpelu and A. D. Agboola, Int. J. Theor. Phys. 47(2008)1039.
\bibitem{ADAMS2}B. G. Adams, J. Cizek and J. Paldus, Adv. Quant. Chem. Vol. 19 (1987)1. Reprinted in reference \cite{BARUT}.
\bibitem{ADAMS} B. G. Adams, Algebraic Approach to Simple Quantum Systems, Springer-Verlag, Berlin, 1994.
\bibitem{GUR}Y. Gur and A. Mann, Phys. At. Nucl. 68(2005)1700.

\end{thebibliography}
\end{document}